# PHOTOLUMINESCENCE IN THE TEMPERATURE RANGE 20–230 K OF C60 FULLERITE DOPED WITH NITROGEN MOLECULES

P.V. Zinoviev and V. N. Zoryansky.

*Verkin Institute for Low Temperature Physics and Engineering of the National Academy of Sciences of Ukraine, Kharkiv 61103, Ukraine*

E-mail: zinoviev@ilt.kharkov.ua

The optical properties of $C_{60}$ single crystals, intercalated with nitrogen molecules, were investigated by the spectral-luminescence method in the temperature range 20–230 K. The saturation was carried out under a pressure of 30 atm. of $N_2$ at various temperatures from 200 to 550 °C. For the $C_{60} + N_2$ system, the presence of a temperature boundary of the adsorption crossover of about 420 °C was established (transition from the diffusion mechanism of intercalation - physisorption, to chemical interaction - chemisorption). The temperature dependence of the integrated radiation intensity of a new nitrogen-containing substance based on $C_{60}$ fullerite has been recorded for the first time. Quenching of photoluminescence at low temperatures was found. The observed new effect of low-temperature quenching of photoluminescence is explained by the appearance in the process of chemical interaction between the $N_2$ impurity molecules and the $C_{60}$ matrix of effective exciton trapping centers and nonradiative deactivation of electronic excitation.

Fullerite $C_{60}$ belongs to the class of molecular crystals. Due to the peculiarities of the crystal structure, molecular crystals, both simple and complex, often demonstrate adsorption [1, 2], polymerization [3, 4], nonlinear optical [5 - 7] and other unique physical properties. The results of studies of various physical properties of molecular crystals are of great importance for fundamental and applied problems of solid state physics. Unlike classical molecular crystals (where the molecules are rather rigidly fixed in the crystal lattice), the $C_{60}$ molecules in fullerite crystals can perform rotational motion. The mutual orientational arrangement of neighboring molecules significantly affects the transfer of electronic excitations in the crystal lattice and significantly depends on the presence of impurities in intermolecular voids. This makes fullerite C60 a convenient material for studying lattice dynamics by spectral-luminescent methods under the influence of various intercalants.

For efficient transport of excitons to emission centers in $C_{60}$ crystals, the coherence of states on two neighboring molecules involved in the tunneling exciton hopping is important, which is observed at their definite mutual orientation. This condition is well satisfied in the low-temperature (glassy) phase of pure fullerite up to the temperature of orientational ordering (glass transition) $T_g$, in which the pentagon orientation of $C_{60}$ molecules prevails. As the temperature rises above $T_g$, the probability of reorientation of molecules increases, which inevitably leads to a violation of the coherence of states, a decrease in the exciton mean free path, and, as a consequence, to an increase in the probability of non-radiative deactivation of the excitation. For this reason in pure $C_{60}$ the integral radiation intensity $I\,(T)$ in the low-temperature phase is maximal and practically does not depend on temperature up to $T_g$. Above the glass transition temperature $T_g$, a rather sharp decrease in the radiation intensity occurs, which progresses with temperature increase [8, 9]. The main contribution to the decrease in $I\,(T)$ at $T > T_g$ is made by the luminescence of localized states ("deep X-traps"), the concentration of which is determined, among other things, by the degree and character of filling the intermolecular voids with intermolecular impurity particles. As a rule, intercalation of $C_{60}$ crystals in the mode of physical sorption by simple molecules (including $N_2$) [10 - 16] increases the lattice volume and lowers the temperatures of the orientational ($T_c$) and glass ($T_g$) transitions. This is due to the effect of intracrystalline "negative pressure", which makes it possible to significantly facilitate the reorientation of $C_{60}$ molecules. With an increase of the saturation temperature, the diffusion mechanism of sorption of a molecular impurity into the fullerite lattice sometimes undergoes a transition to a chemical interaction between an impurity and the matrix (chemisorption) and an adsorption crossover is observed.

In this article, we present the results of studies of the temperature behavior of the photoluminescent properties of $C_{60}$ fullerite crystals doped with $N_2$, saturated under the conditions of chemical sorption of nitrogen molecules from the gas phase. The temperature dependences of the integral radiation intensity of a new nitrogen-containing chemical compound have been established for the first time.

For experimental study of the temperature behavior of the photoluminescence of the $C_{60}$-$N_2$ complex, we used a polycrystalline sample in the form of a powder with a granule size of about 0.5 mm and a purity of at least 99.9%. To purify it from residual atmospheric gases before saturation, the fullerite was preliminarily held in a dynamic vacuum of $10^{-3}$ mm Hg at a temperature of 300 °C for 48 hours. After such annealing before saturation with nitrogen, the photoluminescence spectra were recorded, which corresponded to the spectra of pure crystalline fullerite $C_{60}$ [9] practically without impurities. In accordance with the X-ray diffraction data on the change of the lattice parameter $a$ during the saturation of $C_{60}$ with $N_2$ molecules [17], the intercalation mode was chosen corresponding to the mechanism of chemical sorption. The experimental dependence of the change of the lattice parameter $a$ with temperature during the saturation of fullerite $C_{60}$ with gaseous $N_2$ is shown in Fig. 1.

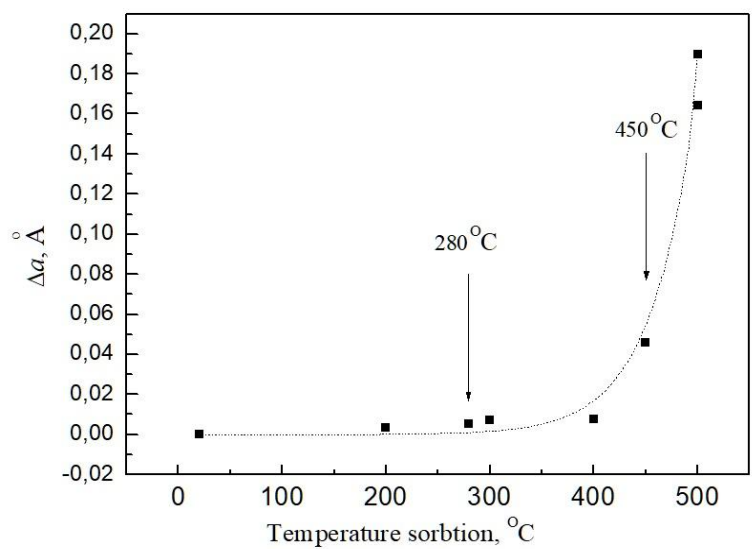

Fig. 1. The dependence of the lattice parameter $a$ of the studied samples of $C_{60}$ on the temperature $T_s$ of saturation with gaseous $N_2$ at a pressure of 30 atm on the base of experimentally obtained data [17].

The sample studied was saturated in the cell at the temperature of 450 °C for 200 hours at a pressure of 30 atm in a nitrogen atmosphere. Such conditions correspond to a region of rapid growth of the lattice parameter a of fullerite $C_{60}$ above the temperature of the adsorption crossover of 420 °C (a characteristic break on the curve).

The photoluminescence (PL) of $C_{60}$ crystals was recorded "for reflection" in the spectral region 1.2-1.85 eV (1033-670 nm) with a spectral resolution of 2 nm using an MDR-2 high-aperture diffraction monochromator and a cooled FEU-62 photomultiplier in the photon counting mode. The cryogenic part of the experimental setup made it possible to change the temperature of the sample in a wide temperature range (20 - 230 K) and to stabilize it during the experiment with an accuracy of 0.5 K for the input power density W≤1 mW/mm$^2$. Such a limitation on the excitation power was introduced to prevent undesirable photostimulated processes in the near-surface layers of the studied polycrystalline samples.

Figure 2c illustrates the temperature dependences of the integral radiation intensities of $C_{60}$ fullerite intercalated with nitrogen at 450 °C and a pressure of 30 atm for 200 h.

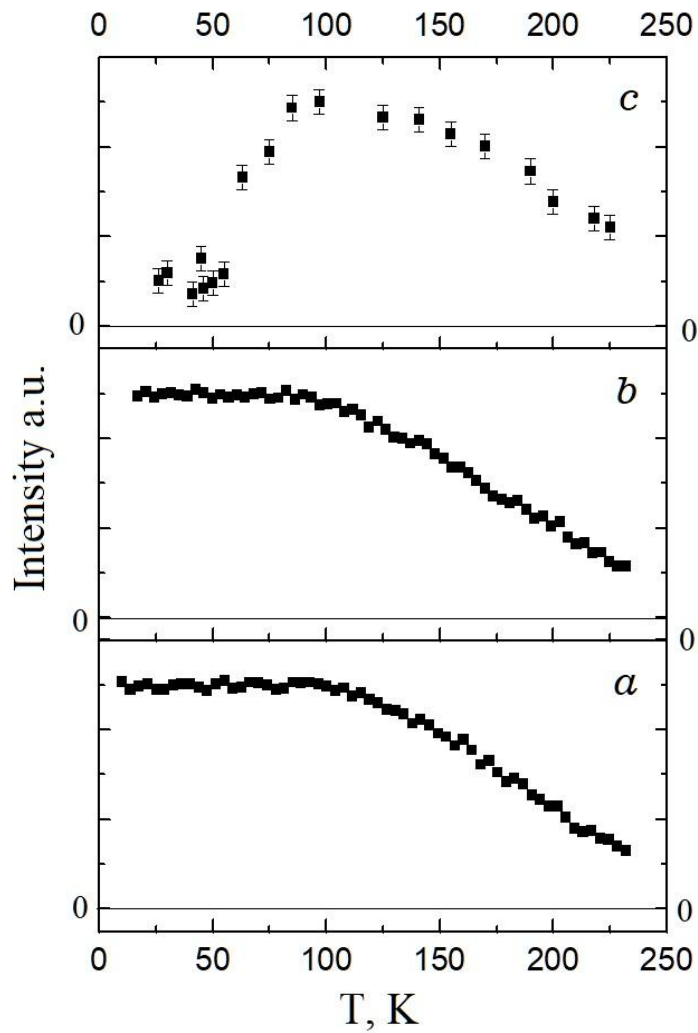

Fig. 2. Temperature dependences of the integrated radiation intensities of the $C_{60}$ fullerite, normalized to the corresponding values at T = 25 K: (a) - pure; (b) - intercalated $N_2$ at 280 °C and a pressure of 30 atm for 150 hours (physical sorption); (c) - intercalated $N_2$ at 450 °C and a pressure of 30 atm for 200 hours (chemical sorption). All curves are measured in the warming mode.

Unlike similar dependences for $C_{60}$, both pure and saturated in the physisorption mode [18], here the points on experimental curves were obtained by processing the photoluminescence spectra of the $C_{60}$-$N_2$ system, recorded at the corresponding temperatures. The study was carried out in the temperature range 20-300 K in the modes of sequential temperature decrease (cooling) and increase (warming). The temperature dependence of the integral radiation intensity in the warming mode can be divided into three stages: a) - low temperatures 25-50 K - consistently low intensity; b) - temperatures 50-100 K - rapid growth of intensity; c) - temperatures 100-230 K - a smooth decrease in the integral radiation intensity. The same stages in the reverse order are observed in the cooling mode. The characteristics of this curve differ significantly from the previously obtained results for pure $C_{60}$ and $N_2$ interstitial solutions (Fig. 2a,b) [18]. As mentioned above, a constant value of the quantum yield from low temperatures to the glass transition temperature and the subsequent decrease in intensity are characteristic of all previously studied two-component systems based on $C_{60}$, obtained by the diffusion mechanism of sorption. Despite the fact that the rotational dynamics of $C_{60}$ molecules in doped fullerite crystals differs from that of pure $C_{60}$, the temperature dependence of the integral photoluminescence intensity always retains a certain character in the entire investigated temperature range. A detailed analysis of the crystal lattice dynamics in terms of the coherent transport of electronic excitations, as well as the mechanisms of the influence of impurities on it, are given in [8].

A significant difference in the behavior of the temperature dependence of the integrated radiation intensity of the samples under study indicates the presence of additional mechanisms of transport of electronic excitations, which are characteristic of the new nitrogen-containing substance based on fullerite $C_{60}$. The low intensity of photoluminescence in the low-temperature region can be caused by the appearance of complex molecular structures in the process of nitriding of fullerene molecules, which play the role of quenching centers with a certain activation energy and a long lifetime. Apparently, such formations at low temperatures are most likely capable of capturing mobile Frenkel excitons with subsequent deactivation of the excitation. Naturally, with increasing temperature, the energy contribution of the vibrational states of these centers increases and, consequently, the probability of the release of localized electronic excitation increases. This, most likely, determines the main mechanism of the increase in the radiation intensity in the 50-100 K region, when the excitons are already able to leave the potential well and migrate through the crystal, reaching the emission centers. With a further increase in temperature, a smooth decrease in the luminescence intensity is observed, the mechanism of which is probably similar to that for pure $C_{60}$ or interstitial solutions [8]. A joint analysis of the results of X-ray diffraction [17], spectral-luminescence studies [19] and studies of the temperature behavior of the photoluminescent properties of fullerite $C_{60}$ doped with $N_2$ indicates the presence of an adsorption crossover for the $C_{60}$-$N_2$ system when the intercalation temperature rises above 420 ºC, with a change in the diffusion mechanism of interaction to the chemical one.

Thus, in this article, on the basis of studies of the temperature behavior of the photoluminescence of the $C_{60}$-$N_2$ system, for the first time, the phenomenon of low-temperature quenching of radiation was discovered for the case of saturation of fullerite with nitrogen in the chemisorption mode (saturation temperature above 420 °C). This unambiguously indicates the presence of a new nitrogen-containing compound in the mixture. The stably low quantum yield of luminescence in the temperature range of 25-50 K can be associated with the formation of new molecular complexes arising in the process of chemical interaction of fullerene and nitrogen molecules. Most likely, the multicomponent system obtained on the basis of $C_{60}$ fullerite contains particles that play the role of highly efficient exciton trapping centers with a sufficient lifetime for nonradiative deactivation of electronic excitation.